# J/$\psi$ and $\psi$(2S) production in p-Pb collisions at $\sqrt{s_{NN}} = 5.02$ TeV with ALICE at the LHC

M. Leoncino for the ALICE Collaboration

*INFN and University, Torino, Italy*

**Summary.** — The ALICE Collaboration has studied the inclusive J/$\psi$ and $\psi$(2S) production in p-Pb collisions at $\sqrt{s_{NN}} = 5.02$ TeV, at the CERN LHC. The J/$\psi$ measurement is performed in the $\mu^+\mu^-$ and in the $e^+e^-$ decay channels, down to zero $p_T$. The results are in fair agreement with theoretical predictions. The $\psi$(2S) measurement has also been performed. In particular, a smaller $\psi$(2S) nuclear modification factor, with respect to the J/$\psi$ one, has been observed.

PACS `25.75.-q` – Relativistic heavy-ion collisions.
PACS `12.38.Mh` – Quark-gluon plasma.
PACS `14.40.Pq` – Heavy quarkonia.

## 1. – Introduction

The suppression of charmonia, bound states of $c$ and $\bar{c}$ quarks, is considered a clean signature of Quark-Gluon Plasma (QGP) formation in heavy-ion collisions [1]. In addition to the color screening mechanism, other effects may contribute to the charmonium production in Pb-Pb collisions. In particular, one can expect a recombination of $c$ and $\bar{c}$ pairs from the medium (favoured by the large $c\bar{c}$ multiplicity typical at the LHC energies [2]). Furthermore, cold nuclear matter (CNM) effects, like shadowing [3,4] and initial state parton energy loss [5], are expected to play a role. It is thus important to study the production of charmonia in proton-nucleus collisions in order to disentangle the suppression contribution related to CNM from the one associated to the formation of a QGP.

## 2. – The ALICE detector and the p-Pb run

The ALICE detector consists of a central barrel dedicated to particle tracking and identification (in the range $|\eta| < 0.9$) and a forward spectrometer used for the detection of muons (in the interval $-4 < \eta < -2.5$). More details about the experimental setup can be found in [6]. The J/$\psi$ resonance is detected both in the dielectron decay channel (using the central barrel detectors) and in the dimuon decay channel (using the forward

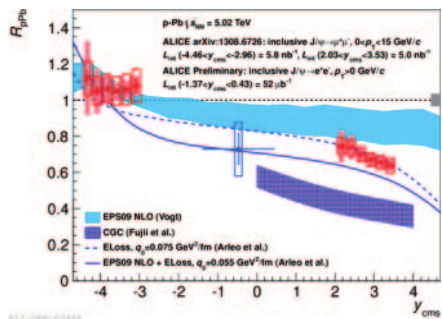

Fig. 1. – Inclusive J/$\psi$ nuclear modification factor as a function of rapidity. Open boxes are uncorrelated systematic uncertainties, filled area indicate partially correlated uncertainties and the gray box at one the global uncertainty due to the $T_{\rm pPb}$. Theoretical models based on shadowing/energy loss [9-11] and CGC [12] are also shown.

muon spectrometer), while, with the present statistics, the $\psi(2S)$ can be studied only in the dimuon decay channel. Due to the energy asymmetry of the LHC beams in p-Pb collisions the nucleon-nucleon center-of-mass system is shifted by $\Delta y = 0.465$ in the direction of the proton beam. Data have been collected in two beam configurations with inverted beam directions, resulting in the following rapidity coverages: $-4.46 < y_{\rm cms} < -2.96$ at backward rapidity, $-1.37 < y_{\rm cms} < 0.46$ at midrapidity and $2.03 < y_{\rm cms} < 3.53$ at forward rapidity.

## 3. – Results

The nuclear effects on J/$\psi$ production in p-Pb collisions are quantified by means of the nuclear modification factor, which is defined by: $R_{\rm pPb} = Y_{\rm J/\psi}/\langle T_{\rm pPb}\rangle \cdot \sigma_{\rm pp}^{\rm J/\psi}$, where $Y_{\rm J/\psi}$ is the efficiency corrected J/$\psi$ yield, $\sigma_{\rm pp}^{\rm J/\psi}$ is the production cross section in pp collisions in the same kinematical range at the same energy and $T_{\rm pPb}$ is the nuclear thickness function estimated through the Glauber model [7]. Since pp data at $\sqrt{s} = 5.02$ TeV are not available, the reference $\sigma_{\rm pp}^{\rm J/\psi}$ is obtained with an interpolation procedure [8]. The results are reported in fig. 1: at mid and forward rapidity, the inclusive J/$\psi$ production is suppressed with respect to that in pp collisions, whereas it is unchanged at backward rapidity. Models containing shadowing and/or energy loss [9-11] are in agreement with ALICE data (within uncertainties), while the CGC-based model [12] overestimates the suppression.

Since the Bjorken x-values in the Pb nucleus in p-Pb collisions at $\sqrt{s_{NN}} = 5.02$ TeV are similar to the ones in Pb-Pb collisions at $\sqrt{s_{NN}} = 2.76$ TeV and assuming a factorization of shadowing effects, an expectation for the $R_{\rm AA}$ based on the $R_{\rm pA}$ can be derived by comparing $R_{\rm pPb}^{forward} \times R_{\rm pPb}^{backward}$ with $R_{\rm PbPb}$ (fig. 2). At low transverse momentum data suggest a contribution from regeneration while, at higher transverse momenta, the suppression contribution starts to be dominant. The $\psi(2S)$ analysis has been performed analogously to the J/$\psi$ one. In fig. 3 (left) the double ratio $[\psi(2S)/{\rm J}/\psi]_{\rm pPb} / [\psi(2S)/{\rm J}/\psi]_{\rm pp}$ is shown as a function of rapidity and is compared with the PHENIX result in d-Au collisions at $\sqrt{s_{NN}} = 0.2$ TeV. ALICE results show a similar trend compared with PHENIX data at midrapidity [13] indicating a suppression in the $\psi(2S)$ production with respect to p-p collisions. This suppression is further investigated in fig. 3 (right) where the $R_{\text{pPb}}^{\psi(2S)}$ is presented as a function of rapidity and is compared to $R_{\text{pPb}}^{J/\psi}$. The $\psi$(2S) is more suppressed with respect to the J/$\psi$. These results are compared to theoretical calculations used for the J/$\psi$ [9-11], which should hold for the $\psi$(2S). The available theoretical predictions do not describe the $\psi$(2S) suppression, indicating that other mechanisms are required to explain it.

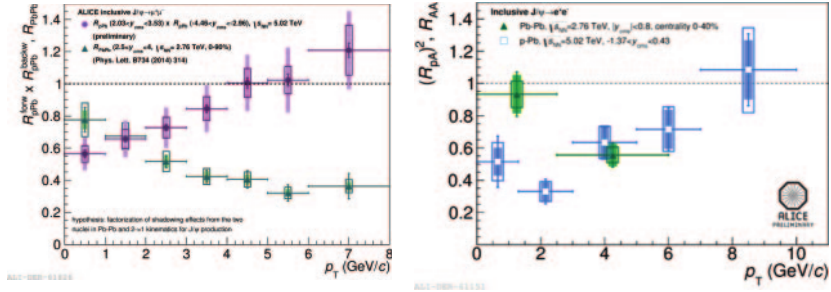

Fig. 2. – In the left plot the J/$\psi$ $R_{\text{pPb}}^{backward} \times R_{\text{pPb}}^{forward}$ is compared to the $R_{\text{PbPb}}^{forward}$. In the right plot the J/$\psi$ is compared $\left(R_{\text{pPb}}^{midrapidity}\right)^2$ with $R_{\text{PbPb}}^{midrapidity}$.

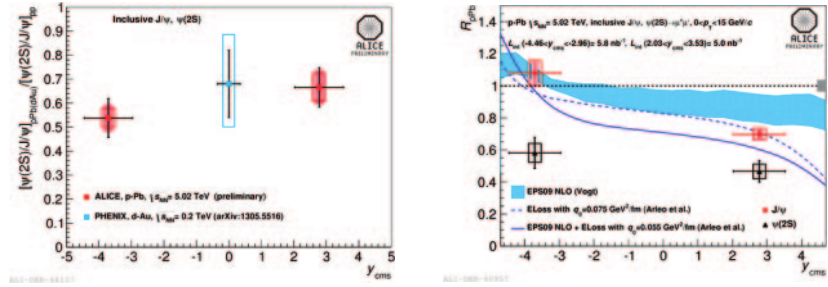

Fig. 3. – Left: the double ratio $[\psi(2S)/J/\psi]_{\text{pPb}} / [\psi(2S)/J/\psi]_{\text{pp}}$. PHENIX data at $\sqrt{s_{NN}} = 0.2$ TeV at midrapidity is also shown. Right: the $R_{\text{pPb}}^{\psi(2S)}$ compared to $R_{\text{pPb}}^{J/\psi}$.